\documentclass{JHEP3}
\newcommand{\be}{\begin{equation}}
\newcommand{\ee}{\end{equation}}
\newcommand{\ben}{\begin{eqnarray}}
\newcommand{\een}{\end{eqnarray}}
\newcommand{\bb}{\bibitem}

\usepackage{graphics}
\usepackage{graphicx}
\usepackage{amsfonts}
\usepackage{amssymb}
\usepackage{epsfig}
\title{Bloch Brane}
\author{D. Bazeia$^a$ and A.R. Gomes$^{a,b}$\\
$^a$Departamento de F\'\i sica, Universidade Federal
da Para\'\i ba, C.P. 5008, 58051-970 Jo\~ao Pessoa PB, Brazil
\\
$^b$Departamento de Ci\^encias Exatas, Centro Federal de Educa\c c\~ao
Tecnol\'ogica do Maranh\~ao, 65025-001 S\~ao Lu\'\i s MA, Brazil}
\abstract{We investigate a system described by two real scalar fields coupled
with gravity in (4, 1) dimensions in warped spacetime involving one
extra dimension. The results show that the parameter which controls the way
the two scalar fields interact induces the appearence of thick brane which
engenders internal structure, driving the energy density to localize inside
the brane in a very specific way.
}
\begin{document}
\maketitle
\section{Introduction}

This work deals with scalar fields coupled to gravity in warped spacetime in (4,1)
dimensions. We focus mainly on generation of thick branes, their stability and
internal structure, and our investigations follow several recent works on the
subject \cite{add,aadd,rs,rs1,gw,df,ce,ceg,g,c,gp,mps,bbn,es,bfg}.

The brane world scenario that we investigate is described by scalar matter coupled
to gravity. Thus, in flat spacetime the scalar
matter model reduces to a system of scalar fields which supports
interesting domain wall solutions. As one knows, domain walls are extended
structures which may be seem as interfaces separating two distinct
phases of the system. They appear in several branches of nonlinear science,
in particular in high energy physics \cite{vs} and in condensed matter \cite{e,w}.
In condensed matter, for instance, domain walls may spring
in several different ways, depending on the richness of the systems
under consideration. An interesting situation occurs in ferromagnetic systems,
as reported for instance in \cite{e,w}, in which one finds walls of both Ising
and Bloch type. Ising walls are simpler interfaces, having no internal
structure. However, Bloch walls are interfaces that engender internal structure,
and for this reason they may be seen as chiral interfaces. The chiral feature
of Bloch walls was first explored in Ref.~{\cite{cou}}; other recent
investigations on Bloch walls in parametrically driven systems may be
found in {\cite{var}}, and in references therein.

In high energy physics, domain walls usually appear in models described
by real scalar fields \cite{vs}. In particular, in Ref.~{\cite{bs}} one
introduces a simple model, described by two real scalar fields, which support
both Ising and Bloch walls -- see, e.g., Ref.~{\cite{bbb,bb,bb1}} for further
details on the two-field model, and on Ising and Bloch walls. The approach
introduced in Ref.~{\cite{bs}} engenders an important issue, which concerns
the Bogomol'nyi bound \cite{b}. In Ref.~{\cite{bs,bbb,bb,bb1}} it was shown
that this bound can be obtained in two-field models defined by the potential
\be\label{p}
V(\phi,\chi)=\frac12\left(\frac{\partial W}{\partial\phi}\right)^2+
\frac12\left(\frac{\partial W}{\partial\chi}\right)^2
\ee
where $W=W(\phi,\chi)$ is a smooth function of the two fields $\phi$ and $\chi.$ 
The two-field model is described by {\cite{bs}}
\be\label{wr}
W_r(\phi,\chi)=\phi-\frac13\phi^3-r\phi\chi^2
\ee
where $r$ is a real parameter, and we are using dimensionless fields and spacetime
variables. The richness of the two-field model was further explored in other
contexts in Refs.~\cite{bnrt,sv,spain}; for instance, in Ref.~{\cite{spain}} the
complete set of solutions for the Bogomol'nyi-Prasad-Sommerfield (BPS)
states \cite{b,ps} of the model was found. We notice that in the model
described by $W_r(\phi,\chi),$ the case $r=0$ lead us back to the standard
$\phi^4$ model.

More recently, domain walls have been used in high energy physics as
seeds for generation of branes in scenarios where scalar fields
couple with gravity in warped spacetime. In particular, in Ref.~{\cite{bfg}}
we have used a model described by a single real scalar field to offer a scenario
where the brane engenders internal structure. The model used in \cite{bfg} was first
introduced in Ref.~{\cite{bmm}}, and it depends on a real parameter, which
generalizes the standard $\phi^4$ model, changing the way the scalar field
self-interacts. It offers an interesting alternative to models which use
two real scalar fields \cite{c,es} as seeds for brane formation in
warped spacetime in $(4,1)$ dimensions. The model is defined by \cite{bmm}
\be\label{wp}
W_p(\phi)=\frac{p}{2p-1}\phi^{(2p-1)/p}-
\frac{p}{2p+1}\phi^{(2p+1)/p}
\ee
where $p$ is an odd integer, $p=1,3,5,...$ The case $p=1$ reproduces the standard
$\phi^4$ model. The models defined by $W_p(\phi)$ support BPS solutions of the form 
\be
\phi_p(x)=\tanh^p(x/p)
\ee
The case $p=1$ reproduces the standard kink, but for $p=3,5,...$ one gets
new solutions, which very much resemble two-kink solutions \cite{bmm}.
In the above model, however, the scalar field self-interacts in an unconventional
manner. For this reason, in the present work we return to the issue concerning
thick branes that engender internal structure in models which support conventional
self-interactions.

As one knows \cite{mb}, the presence of intricate self-interactions in general
responds to thicken the brane \cite{g,bfg}. Thus, in the present work we offer
a way to trade the intricate self-interactions contained in the above model
(\ref{wp}), to a more conventional model described by two real scalar fields,
with the potential defined by Eqs.~(\ref{p}) and (\ref{wr}). Within this context,
it is interesting to see that the model defined by (\ref{wr}) is an extension of
the standard $\phi^4$ model to the case of two real scalar fields, while the model
defined by (\ref{wp}) extends the  $\phi^4$ model to the case of more intricate
self-interactions. To implement the investigation, we split the work in the several
sections that follow. In the next Sect.~{\ref{bb}} we provide a complete
investigation, which shows how both the Ising and Bloch walls translate to stable
branes in scenarios where the scalar fields couple to gravity in warped spacetime.
In Sect.~{\ref{is}} we show how the second field required to implement the Bloch
brane scenario contributes to engender internal structure to the brane solution.
Finally, in Sect.~{\ref{cc}} we present our comments and conclusions.

\section{Structure formation}
\label{bb}

In this Section we first deal with domain walls in flat spacetime, which
includes the basic issues one needs to investigate Bloch brane in curved spacetime.
For clarity, we split the subject in two subsections, one dealing with Bloch walls,
and the other with Bloch branes. We start the investigation with Bloch walls in
flat spacetime in the subsection that follows. 

\subsection{Bloch wall}

We first work in flat spacetime in $(3,1)$ dimensions. We consider the action
\be
S=\int d^4x \biggl[\frac12\partial_\mu\phi\partial^\mu\phi+
\frac12\partial_\mu\chi\partial^\mu\chi-V(\phi,\chi)\biggr]
\ee
for $\mu,\nu=0,1,2,3,$ with metric $ds^2=\eta_{\mu\nu}dx^{\mu}dx^{\nu},$ where
${\rm diag}(\eta_{\mu\nu})=(1,-1,-1,-1).$ We consider the potential
\be
V(\phi,\chi)=\frac12(1-\phi^2-r\chi^2)^2+2r^2\phi^2\chi^2
\ee
which is defined by Eq.~(\ref{p}), with $W=W_r(\phi,\chi)$ given by Eq.~(\ref{wr}).
This potential has minima at the points $(\pm1,0)$ and $(0,\pm 1/\sqrt{r})$
for $r$ positive.

The equations of motion for static solutions $\phi=\phi(x)$ and
$\chi=\chi(x)$ that follow from the above model are solved by the
first-order differential equations
\ben\label{dphidx}
\frac{d\phi}{dx}&=&1-\phi^2-r\chi^2
\\
\label{dchidx}
\frac{d\chi}{dx}&=&-2r\phi\chi
\een
We solve these equations using
two distinct orbits connecting the minima $(\pm1,0)$: first, we consider
$\chi=0,$ which identifies a straight line segment joinning the minima
$(\pm1,0).$ In this case we get the one-field solution $\phi(x)=\tanh(x).$
Second, we consider the orbit
\be
\phi^2+\frac{r}{1-2r}\chi^2=1
\ee
for $r\in(0,1/2).$ This orbit identifies elliptical segments connecting the minima
$(\pm1,0).$ In this case we get the two-field solutions
\ben\label{phi}
\phi(x)&=&\tanh(2rx)
\\ 
\label{chi}
\chi(x)&=&\pm\sqrt{\frac1r-2\,}\;{\rm sech}(2rx)
\een
We notice that the limit $r\to1/2$ changes the two-field solution to the
one-field solution. The two-field solutions represent Bloch walls,
while the one-field solution is the Ising wall. We make this point clearer
by recalling that the Ising wall is described by a straight line segment
joinning the minima $(\pm1,0)$ in the $(\phi,\chi)$ plane. For the Bloch walls,
however, the orbits represent elyptical segments connecting the minima $(\pm1,0),$
which may be used to describe chiral interfaces. The above model supports
other solutions, as found for instance in Refs.~{\cite{sv,spain}}.

\subsection{Bloch brane}

We now change to curved spacetime. We consider the action
\be
S_c=\int d^4x\,dy\sqrt{|g|}\Bigl[-\frac14 R+
\frac12\partial_a\phi\partial^a\phi+\frac12\partial_a\chi\partial^a\chi-
V(\phi,\chi)\Bigr]
\ee
where $g=\det(g_{ab})$ and the metric is
\be
ds^2=g_{ab}dx^adx^b=e^{2A}\eta_{\mu\nu}dx^{\mu}dx^{\nu}-dy^2
\ee
where $a,b=0,1,2,3,4,$ and $e^{2A}$ is the warp factor. We are working
in $(4,1)$ spacetime dimensions, using $x^4=y$ to identify the extra dimension.

We suppose that the scalar fields and the warp factor only depend on the extra
coordinate $y$. In this case the equations of motion reduce to the simpler form
\ben
\phi^{\prime\prime}+4A^\prime\phi^\prime&=& \frac{\partial
V(\phi,\chi)}{\partial\phi}
\\
\chi^{\prime\prime}+4A^\prime\chi^\prime&=& \frac{\partial
V(\phi,\chi)}{\partial\chi}
\\
A^{\prime\prime}&=&-\frac23\,\left(\phi^{\prime2}+\chi^{\prime2}\right)
\\
A^{\prime2}&=&\frac16\left(\phi^{\prime2} +
\chi^{\prime2}\right)-\frac13 V(\phi,\chi)
\een
where prime stands for derivative with respect to $y$. Our model is given by
\be\label{gpot}
V_c(\phi,\chi)=\frac18 \left[\left(\frac{\partial
W_c}{\partial\phi}\right)^2 + \left(\frac{\partial
W_c}{\partial\chi}\right)^2 \right]-\frac13 W^2_c
\ee
where $W_c=2W_r,$ that is, it is twice the function we have used in flat
spacetime; see, e.g., Eq.~(\ref{wr}).

The above potential leads to the following first-order differential equations,
which also solve the equations of motion \cite{df}
\be
\phi^{\prime}=\frac12\,\frac{\partial W_c}{\partial\phi}
\ee
\be
\chi^{\prime}=\frac12\,\frac{\partial W_c}{\partial\chi}
\ee
and
\be\label{Aprime}
A^\prime=-\frac13\,W_c
\ee

We solve the first-order equations for $\phi$ and $\chi$ to get the same
results we have already obtained in flat spacetime, given by Eqs.~(\ref{phi})
and (\ref{chi}), now changing $x\to y,$ to represent the extra dimension.
They are depicted in Fig.~[1] for several values of the parameter $r$. We see
that the width of these solutions depends on $1/r$, so it increases for
decreasing $r.$

\begin{figure}[!ht]
\includegraphics[{angle=270,width=7.5cm}]{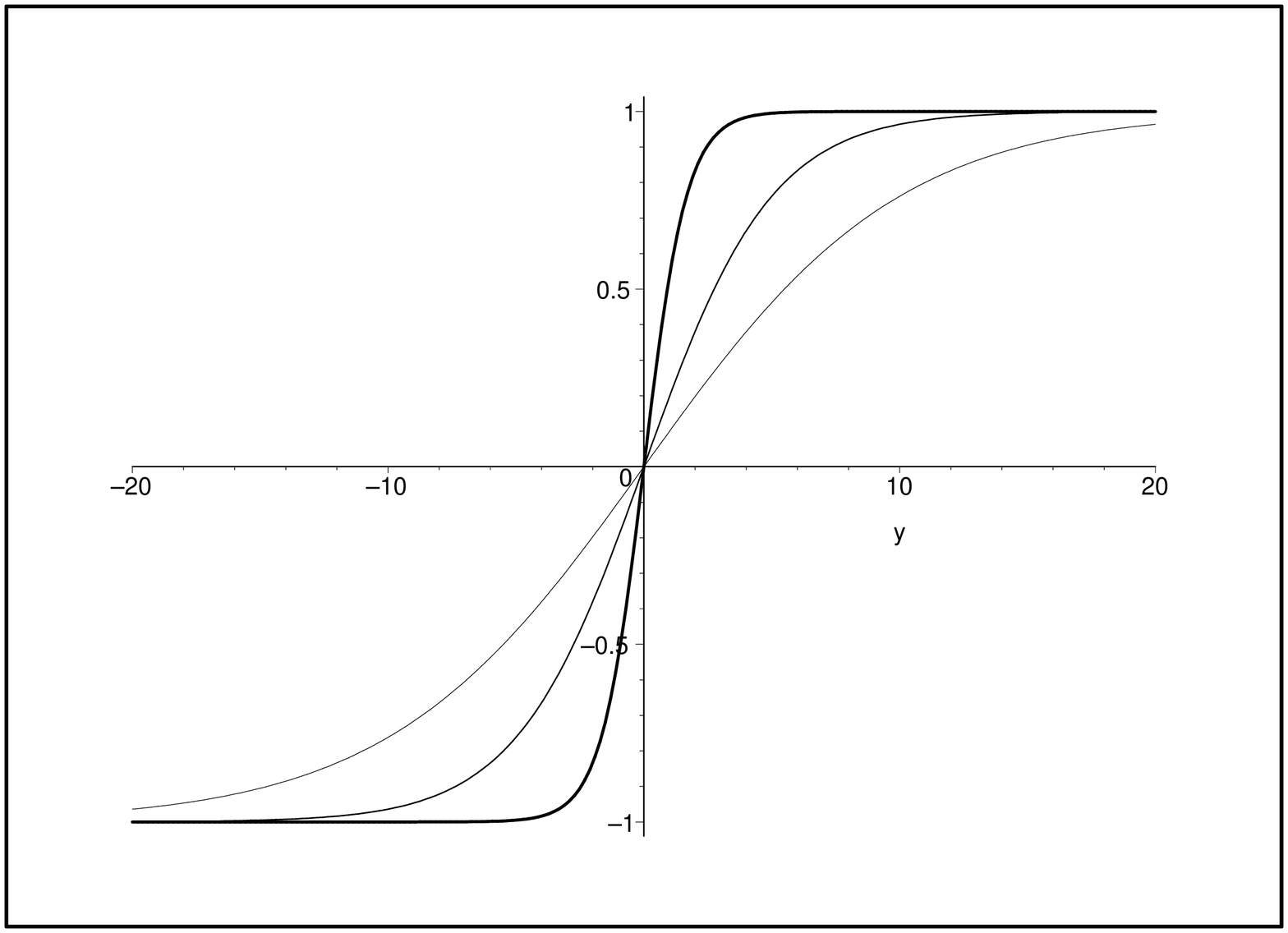}
\includegraphics[{angle=270,width=7.5cm}]{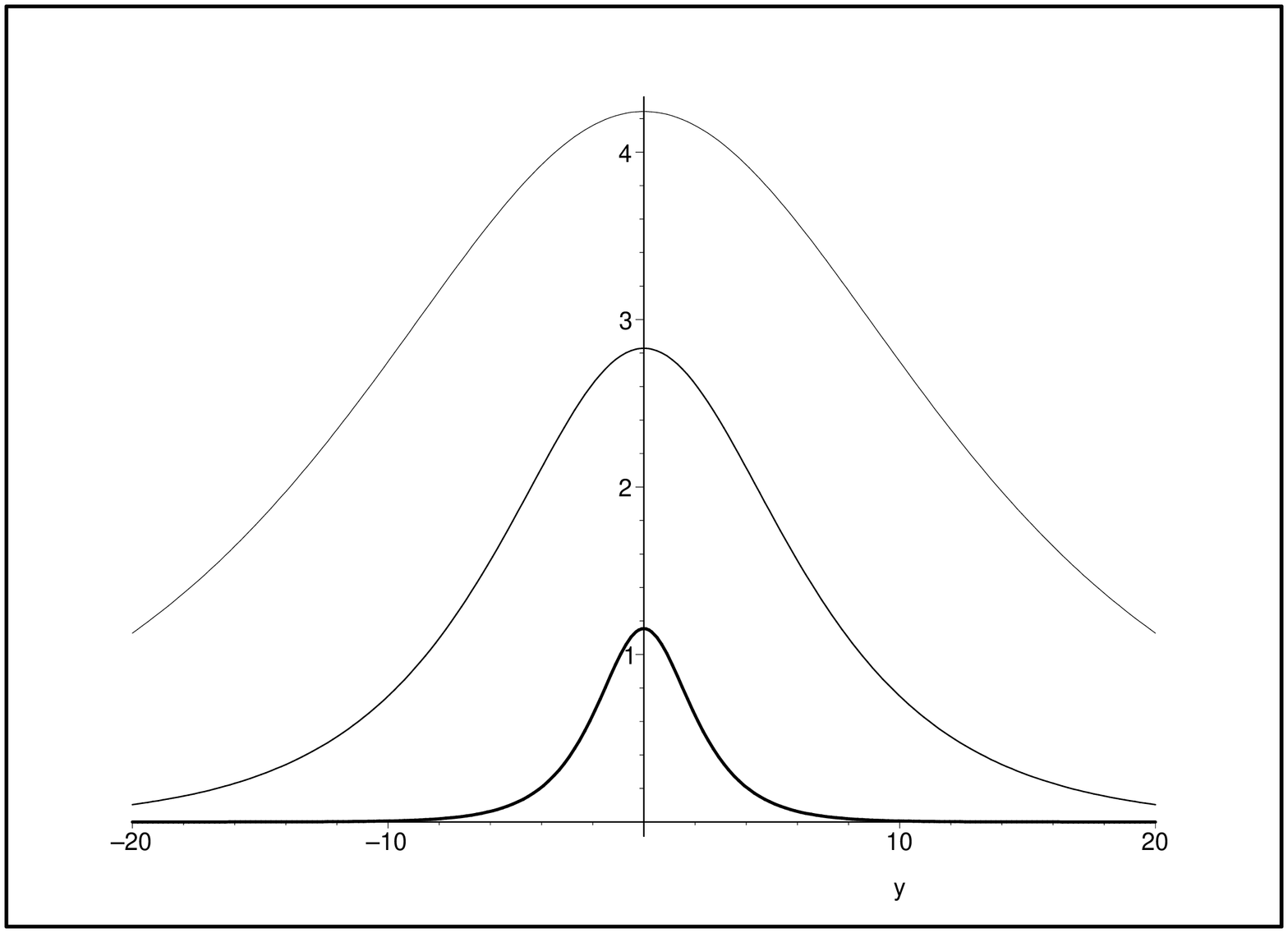}
\caption{Plots of the solutions $\phi(y)$ (right) and $\chi(y)$ (left)
for $r=0.05,0.1\;{\rm and}\;0.3$ respectively.  Here and in the other
figures the thickness of the lines increases with increasing $r.$}
\end{figure}

We now turn attention to the behavior of the warp factor. We solve
Eq.~(\ref{Aprime}) to get
\ben\label{warp}
A(y)\!\!&=&\!\!\frac1{9r}\Bigl[(1-3r)\tanh^2(2ry)-2\ln\cosh(2ry)\Bigr]
\een
We plot the warp factor $e^{2A(y)}$ for several values of
the parameter $r$ in Fig.~[2]. It characterizes the presence of
thick branes, in accordance with the behavior of the solutions
(\ref{phi}) and (\ref{chi}) for the scalar fields.

\begin{figure}[!ht]
\includegraphics[{angle=270,width=10cm}]{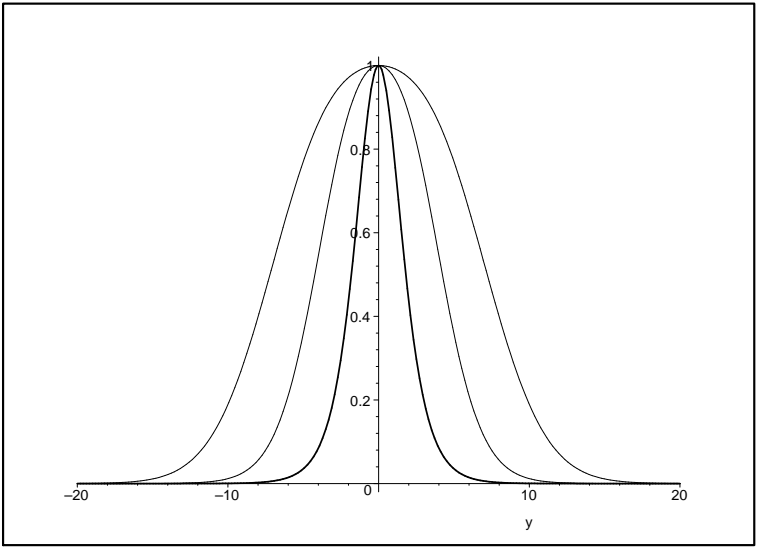}
\caption{Plots of the solutions of the warp factor $\exp[2 A(y)]$ for the values
$r=0.05, 0.1\,{\rm and}\,0.3.$}
\end{figure}

We now consider fluctuations of the metric and scalars.
The perturbed metric is
\be
ds^2=e^{2A(y)}(\eta_{\mu\nu}+\epsilon h_{\mu\nu})dx^\mu
dx^\nu-dy^2
\ee
and we consider $\phi\rightarrow\phi+\epsilon\tilde{\phi}$ and
$\chi\rightarrow\chi+\epsilon\tilde{\chi}$. We are using
$h_{\mu\nu}=h_{\mu\nu}(x,y)$, $\tilde\phi=\tilde\phi(x,y),$
and $\tilde\chi=\tilde\chi(x,y)$ to represent the respective perturbations;
here $x$ represents $(x^0,x^1,x^2,x^3)$. We consider variation of the action with
respect to the scalar fields up to second order in $\epsilon$ to obtain the
equations for the scalar fluctuations. The procedure follows Ref.~{\cite{dwf}}
and we get, for $\tilde\phi$ 
\be
e^{-2A}\Box\tilde\phi -
4A^{\prime}{\tilde\phi}^{\prime}-\tilde\phi^{\prime\prime} +
\tilde\phi\frac{\partial^2V(\phi,\chi)}{\partial\phi^2}
+\tilde\chi\frac{\partial^2V(\phi,\chi)}{\partial\phi\partial\chi}=
\frac12\phi^{\prime}\eta^{\mu\nu}h^{\prime}_{\mu\nu},
\ee
and for $\tilde\chi$
\be
e^{-2A}\Box\tilde\chi -
4A^{\prime}{\tilde\chi}^{\prime}-\tilde\chi^{\prime\prime} +
\tilde\chi\frac{\partial^2V(\phi,\chi)}{\partial\chi^2}+
\tilde\phi\frac{\partial^2V(\phi,\chi)}{\partial\chi\partial\phi}
= \frac12\chi^{\prime}\eta^{\mu\nu}h^{\prime}_{\mu\nu}
\ee
where $\Box=\partial_\mu\partial^\mu.$ Also, we vary the action with respect
to the metric up to second order in $\epsilon$ to obtain
\ben
&&-\frac12\Box h_{\mu\nu}+e^{2A}\biggl(\frac12\partial^2_y+
2A^{\prime}\partial_y\biggr)h_{\mu\nu}
-\frac12\eta^{\lambda\rho}(\partial_{\mu}
\partial_{\nu}h_{\lambda\rho}-\partial_{\mu}\partial_{\lambda}h_{\rho\nu}
-\partial_{\nu}\partial_{\lambda}h_{\rho\mu})\nonumber
\\
&&+\frac12\eta_{\mu\nu}e^{2A}A^{\prime}\partial_y(\eta^{\lambda\rho}h_{\lambda\rho})
+\frac43 e^{2A}\eta_{\mu\nu}\biggl(\tilde\phi\frac{\partial
V(\phi,\chi)}{\partial\phi}+\tilde\chi\frac{\partial
V(\phi,\chi)}{\partial\chi} \biggr)=0
\een
We choose $h_{\mu\nu}$ as transverse and traceless \cite{df}, that is, we change
$h_{\mu\nu}\to{\bar h}_{\mu\nu}=P_{\mu\nu\lambda\rho}h^{\lambda\rho},$
where the projector is given by
\be
P_{\mu\nu\lambda\rho}=\frac12(\pi_{\mu\lambda}\pi_{\nu\rho}+
\pi_{\mu\rho}\pi_{\nu\lambda})-\frac13\pi_{\mu\nu}\pi_{\lambda\rho}
\ee
with
\be
\pi_{\mu\nu}=\eta_{\mu\nu}-
\frac{\partial_\mu\partial_\nu}{\Box}
\ee
In this case the equation for fluctuation
of the metric simplifies to
\be\label{h} {\bar
h}_{\mu\nu}^{\prime\prime}+4\,A^{\prime} \,{\bar
h}_{\mu\nu}^{\prime}=e^{-2A}\,\Box\,{\bar h}_{\mu\nu}
\ee
which describes linearized gravity, in the transverse and traceless sector.
We notice that in this sector, gravity decouples from the matter field.

We use another coordinate, $z$, which is defined by $dz=e^{-A(y)}dy$. Also,
we set
\be
{\bar h}_{\mu\nu}(x,z)=e^{ik\cdot
x}e^{-\frac{3}{2}A(z)}H_{\mu\nu}(z)
\ee
In this case the above
equation (\ref{h}) becomes the Schr\"odinger-like equation
\be\label{se}
-\frac{d^2H_{\mu\nu}}{dz^2}+U_r(z)\,H_{\mu\nu}=k^2\,H_{\mu\nu}
\ee
where the potential is given by
\be
U_r(z)=\frac32\,A^{\prime\prime}(z)+\frac94\,A^{\prime2}(z)
\ee
One can show that the Hamiltonian in eq.~(\ref{se}) leads to
no tachyonic excitations\cite{bfg}. For the zero modes $(k=0)$
we have
\be
H_{\mu\nu}(z)=N_{\mu\nu}e^{3A(z)/2}
\ee
where $N_{\mu\nu}$ is a normalization factor. In Fig.~[3] we
plot the potentials $U_r(z)$ and the corresponding zero modes for
some values of $r$. We see that $U_r(z)$ is a volcano-like potential,
which may lead to the appearance of resonant states in its spectrum.
For this reason, in Fig.~[4] we depict some wave functions for massive modes
in the two regions where $k^2<{\rm max}[U_r(z)]$ and $k^2>{\rm max}[U_r(z)].$
The results show that the wave functions describe motion in the bulk, not bound
to the brane, with no evidence for resonance when $k^2<{\rm max}[U_r(z)]$.

\begin{figure}[!ht]
\includegraphics[{angle=270,width=7.5cm}]{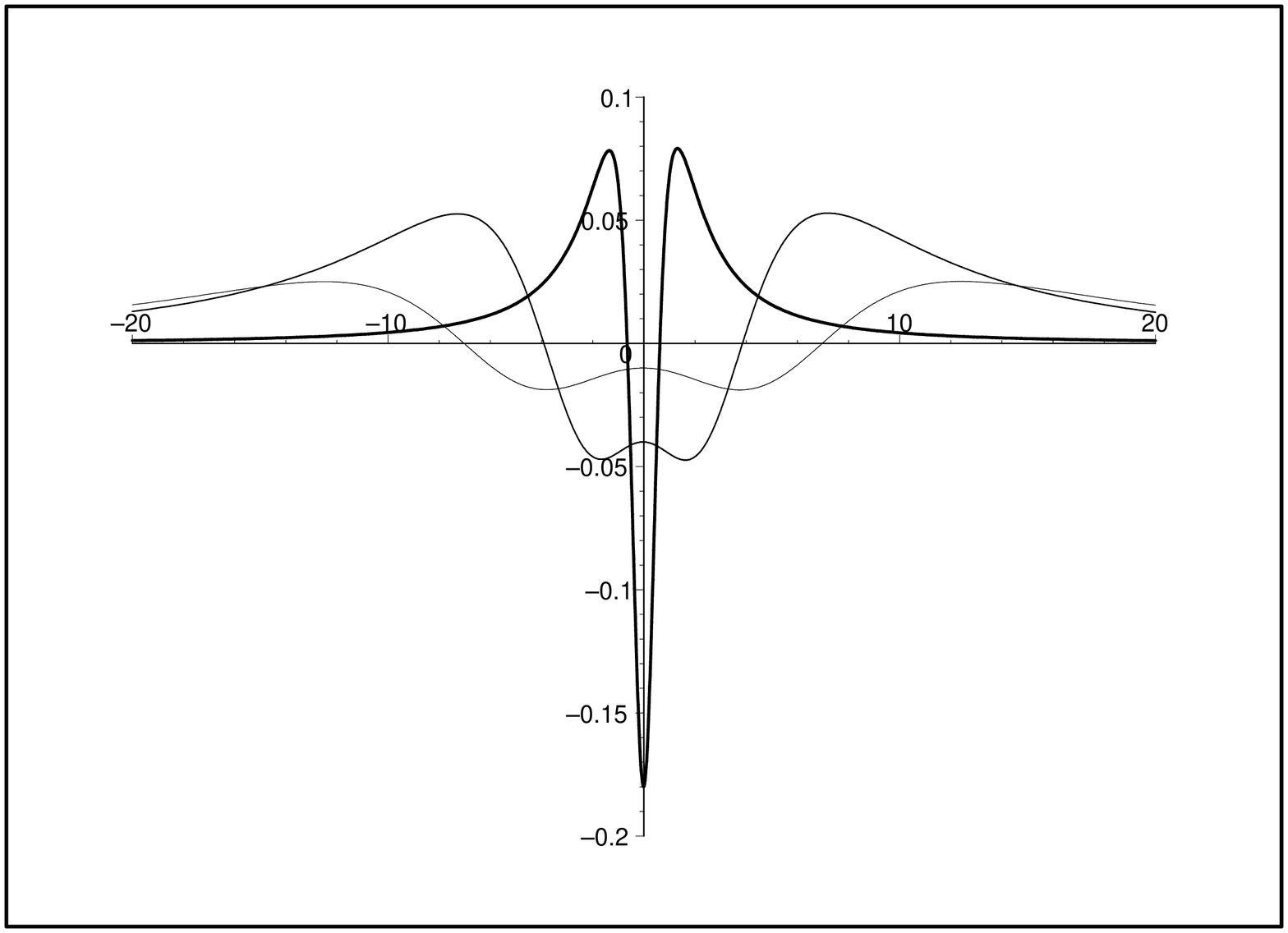}
\includegraphics[{angle=270,width=7.5cm}]{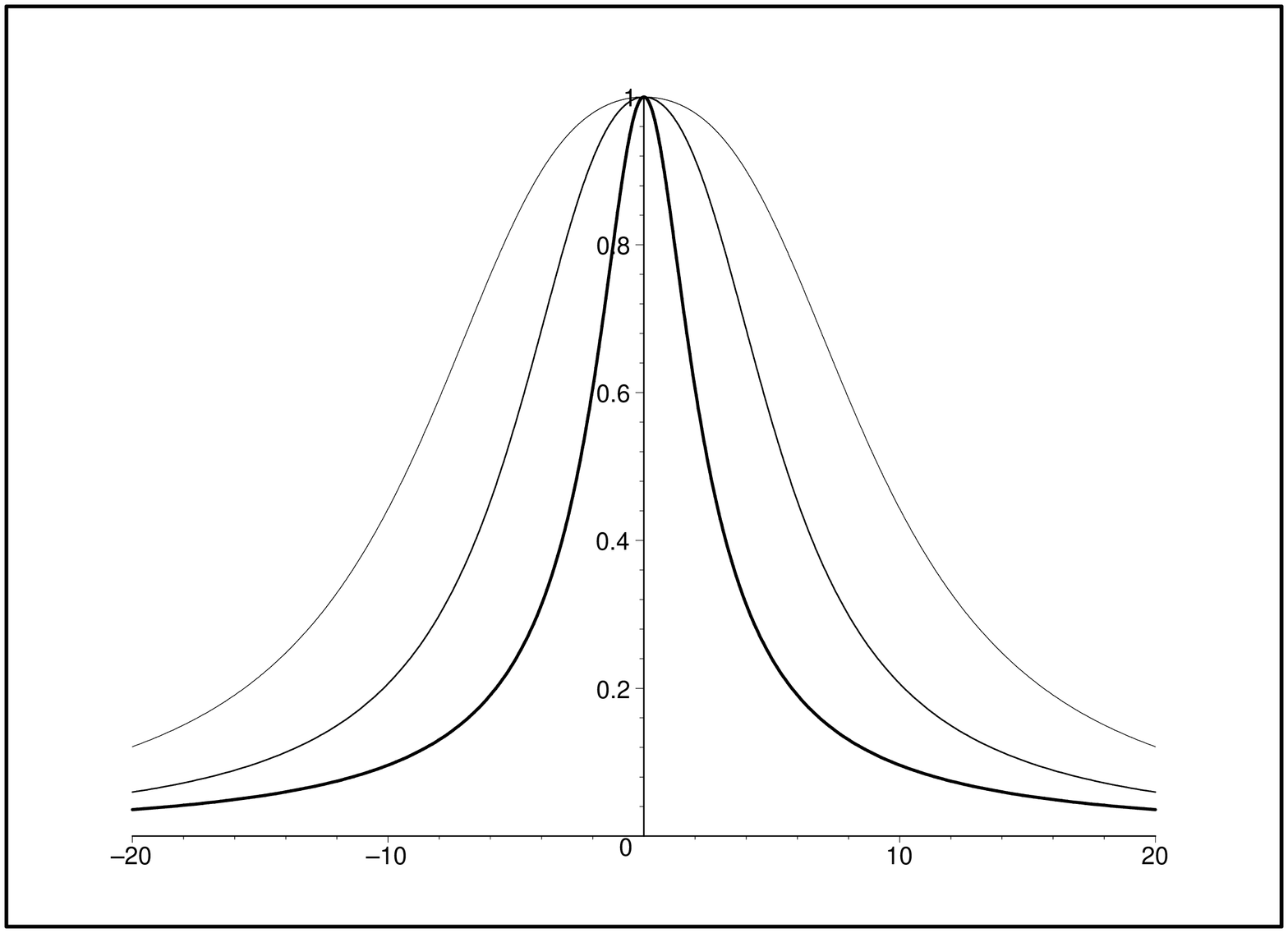}
\caption{Plots of the potentials $U_{0.05}(z), U_{0.1}(z),$
and $(1/2) U_{0.3}(z)$ (right) and the corresponding zero modes (left).}
\end{figure}
\begin{figure}[!ht]
\includegraphics[{angle=270,width=7.5cm}]{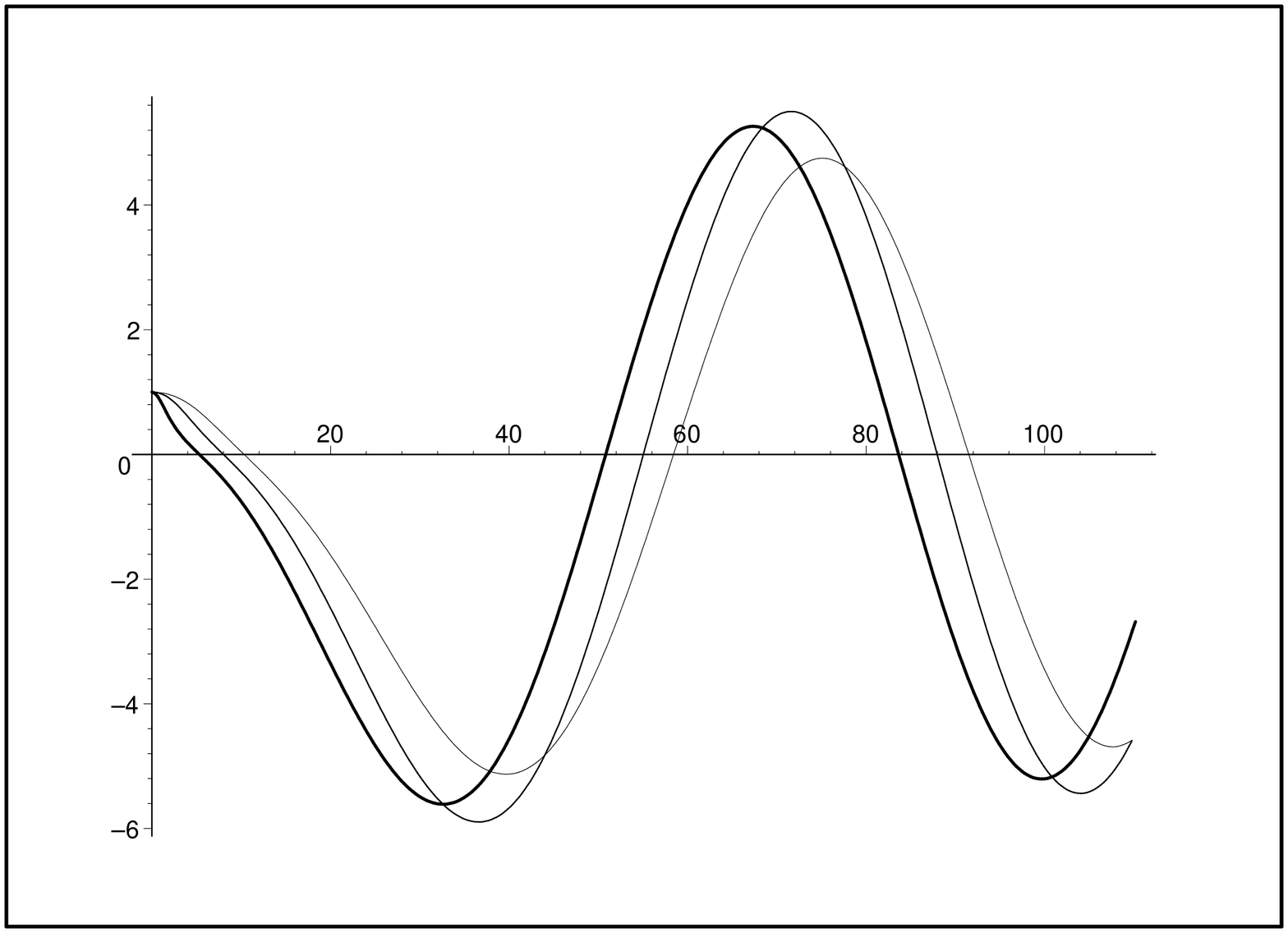}
\includegraphics[{angle=270,width=7.5cm}]{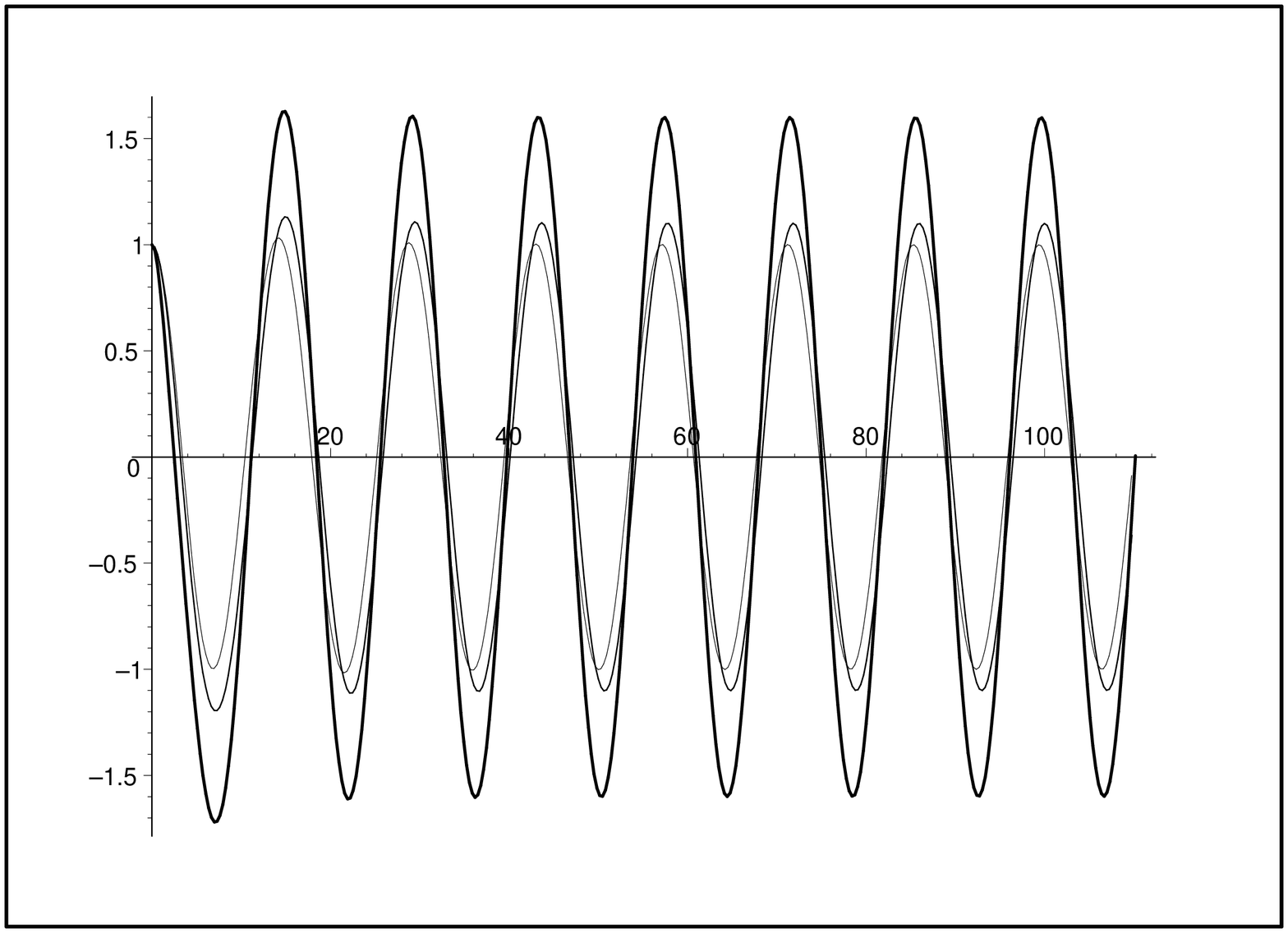}
\caption{Plots of wave functions for nonzero modes for $k^2=0.01<{\rm
max}[U_r(z)]$ (right) and for $k^2=0.2>{\rm max}[U_r(z)]$ (left) for
$r=0.05,0.1,\;{\rm and}\;0.3.$}
\end{figure}

The above calculations show that the Bloch brane is consistent
and cannot be discarded as an alternative to the brane scenario.
More importantly, the Bloch brane scenario nicely leads to the possibility
of brane with internal structure, as we investigate in the next section. 

\section{Internal structure}
\label{is}

To see how the Bloch brane supports internal structure, let us now pay closer
attention to the volcano potential depicted in Fig.~[3]. There we see that it has
a true minimum at $y=0$ for $0.5>r>r_c^*,$ with $r_c^*\approx 0.17.$ However, when
$r$ reaches the interval $r_c^*>r>0,$ the minimum at $y=0$ splits into two separate
minima, indicating the appearence of internal structure. We make this point clearer
with the help of our former work on brane structure \cite{bfg}, from which we
very naturally realise that the behavior of the volcano potential as $r$ diminishes
signals for the opening of an internal structure for the Bloch brane.

To make the reasoning unambiguos, however, let us return to flat spacetime to focus
on the issue in the simpler case firstly. We consider the system of
Eqs.~(\ref{dphidx}) and (\ref{dchidx}) for the flat case to see that it
decouples at $r=0,$ leading to $\phi(y)=\tanh(y)$ in that case.
However, in the limit $r\to0$ the two-field solutions (\ref{phi}) and (\ref{chi})
do not reduce to the former solutions. The reason for this is that the solutions
(\ref{phi}) and (\ref{chi}) are exact solutions for the system of first-order
equations; they are not perturbative around $r=0$, so they are not expected to
reproduce the solutions in the limit of vanishing $r.$ Thus, it is exactly the
nonperturbative behavior of the solutions that induce the appearence of nontrivial
behavior for small values of r, for the specific nonlinearity of the system under
consideration.

To see how this works explicitly, we investigate the energy density of the
nonperturbative two-field solutions in flat spacetime. We have
\be\label{E}
{\cal E}^{\,r}(y)=\frac12\left(\frac{d\phi(y)}{dy}\right)^2+
\frac12\left(\frac{d\chi(y)}{dy}\right)^2
+\frac12\left(\frac{\partial W_r}{\partial\phi}\right)^2+
\frac12\left(\frac{\partial W_r}{\partial\chi}\right)^2
\ee
We use the solutions of the matter fields (\ref{phi}) and  (\ref{chi})
to obtain the energy density, which we plot in Fig.~[5] for several
values of $r$.

\begin{figure}[!ht]
\includegraphics[{angle=270,width=10.0cm}]{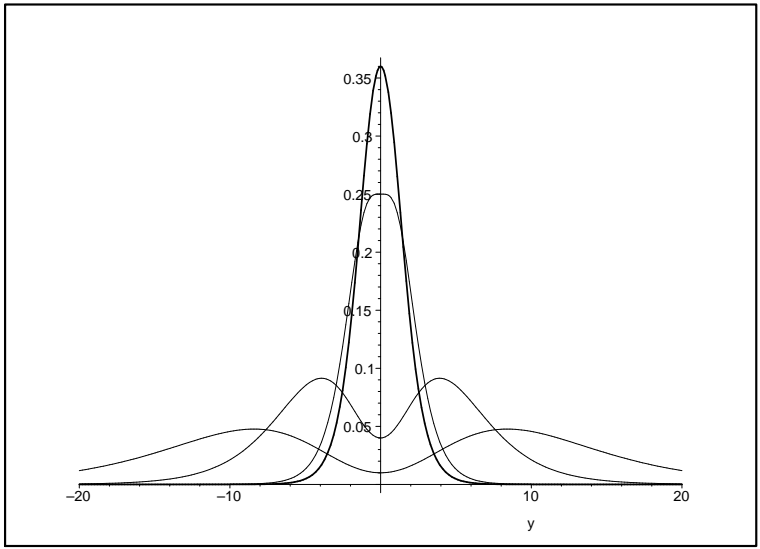}
\caption{The matter energy density in flat spacetime. The plots show
${\cal E}^{\,0.05}(y), {\cal E}^{\,0.1}(y),$ ${\cal E}^{\,0.25}(y)$
and ${\cal E}^{\,0.3}(y).$}
\end{figure}

We see that for $1/2>r\geq1/4$, the energy density has the single-peak
behavior around $y=0,$ as in the standard $\phi^4$ model. However,
for $r^*=1/4$ the plot shows the appearence of a plateau characterized
by the maximum of $d\chi/dy,$ signalling the splitting of the maximum into
two new maxima, as depicted in Fig.~[5]. Here we recognize that it is the
$\chi-$field which responds for changing the behavior of the energy density for
$r$ in the interval $1/4>r>0.$ If we identify the center of the defect with
the maximum of the energy density, we see that for $r^*>r>0$ there appear
two centers, identifying two interfaces which induce the appearence of internal
structure. Moreover, the thickness of the defect increases
for decreasing $r,$ leading to the nice picture where as $r$ decreases from $r=1/2,$
the defect gets thicker and thicker, until $r$ reaches the interval $(1/4,0),$
where two interfaces springs inside the defect. 

We now turn attention to curved spacetime, to the Bloch brane scenario 
investigated above. In this case, the matter energy density has the form
\be\label{Ec}
{\cal E}_c^{\,r}(y)=e^{2A(y)}\biggl[\frac12\left(\frac{d\phi(y)}{dy}\right)^2+
\frac12\left(\frac{d\chi(y)}{dy}\right)^2+V_c(\phi(y),\chi(y))\biggr]
\ee
We use the solutions of the matter fields and the warp factor
(\ref{phi}), (\ref{chi}) and (\ref{warp}) to obtain the matter
energy density as a function of the extra dimension, which we plot
in Fig.~[6] for several values of $r$.

\begin{figure}[!ht]
\includegraphics[{angle=270,width=10.0cm}]{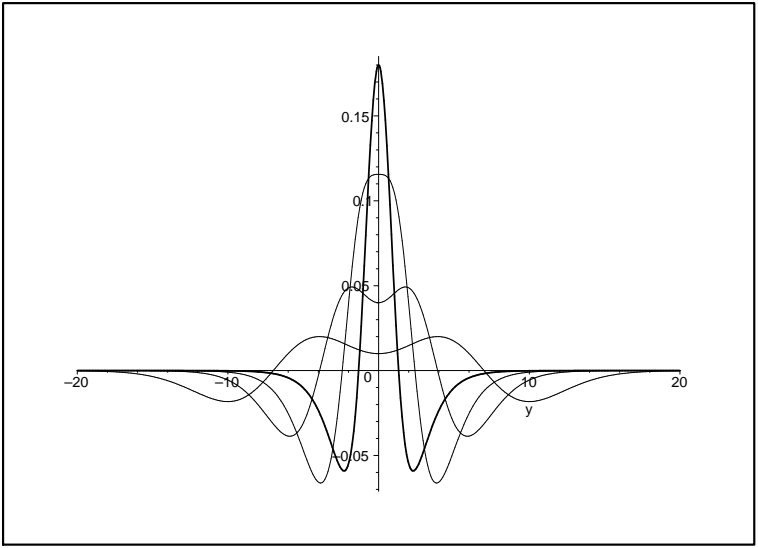}
\caption{The matter energy density in curved spacetime. The plots show
${\cal E}_c^{\,0.05}(y),$\, ${\cal E}_c^{\,0.1}(y)$,\,${\cal E}_c^{\,0.17}(y)$
and $(1/2){\cal E}_c^{\,0.3}(y).$ }
\end{figure}

The investigation shows that for $0.5>r\geq 0.17$ we have the single-peak
behavior in the very inside of the brane. For $r^*_c\approx 0.17$ we
also find a plateau, as we did in the flat case. Here, however, it appear
at the lower value $r^*_c<r^*=0.25,$ due to the coupling to gravity. But
in the region $0.17>r>0$ the behavior for the energy density shows the
appearance of two peaks, as we illustrate in Fig.~[6]. Thus, the coupling
to gravity does not destroy the behavior which induces the presence of internal
structure for the Bloch brane. Moreover, the same effect of increasing the brane
thickness as $r$ goes to zero is observed. Furthermore, we emphasize that for the
same $r$, the thickness of the defect diminishes in the curved case, in comparison
with the flat case. This is compatible with the reasoning that gravity induces
an attraction between the interfaces, as we have already identified in
Ref.~{\cite{bfg}} --- compare Figs.~[5] and [6]. Since the value of $r_c^*$
which induces the appearance of internal structure is lower in curved
spacetime, we can infer that gravity tends to supress the presence of internal
structure in braneword scenarios like the one here investigated.

\section{Further Comments}
\label{cc}

In this work we have investigated the presence of Bloch brane in braneworld
scenarios which appear in models involving real scalar fields coupled to gravity
in $(4,1)$ dimensions in warped spacetime. Our investigations are based on
several recent works on branes \cite{add,aadd,rs,rs1,gw,df,ce,ceg,g}, and it
follows the lines of our former work on brane structure \cite{bfg}.

Former investigations on braneworld models described by scalar fields coupled to
gravity in warped geometry such as the ones in Refs.~\cite{g,bfg} show that the
presence of scalar field with intricate self-interactions directly contributes
to thicken the brane. Furthermore, in \cite{bfg} one shows that
self-interactions may also lead to brane engendering internal structure,
in a way similar to that found in Ref.~{\cite{c}}. This fact has given
the main motivation for the present work: to show how the unconventional
self-interactions introduced in \cite{bfg} can be traded to more conventional
interactions, under the expences of adding more fields to the matter field
contents of the model. The Bloch brane solution here introduced is stable,
and it correctly responds for the presence of nontrivial structure inside the
brane. The addition of the second field allows to trade intricate self-interactions
present in the single-field scenario to the conventional interactions that appear
in the two-field model.

A natural extension of the two-field model, leading to three fields could be
\be
W(\phi,\chi,\rho)=\phi-\frac13\phi^3-r\phi(\chi^2+\rho^2)
\ee
which was already considered before in other contexts \cite{spain1,blw}.
In this case the first-order equations for the matter fields are solved by
\be
\phi(y)=\tanh(2ry)
\ee
and
\ben
\chi(x)&=&\sqrt{\frac1r-2}\cos(\theta){\rm sech}(2ry)
\\
\rho(x)&=&\sqrt{\frac1r-2}\sin(\theta){\rm sech}(2ry)
\een
with $1/2\geq r>0,$ and $2\pi>\theta\geq0.$ The new parameter $\theta$ drives the
elliptical orbits to form an ellipsoid. This extension changes the former scenario
with the inclusion of a new degree of freedom, which may be used to make the second
field complex, allowing for the presence of the global symmetry $U(1),$ which can be
gauged to lead to more realistic models. Other extensions to three-field models could
be done. An interesting line of investigation could follow Ref.~{\cite{bft}}, which
investigates how walls and global and local strings can end on walls. Such
investigations would require the presence of two or more extra dimensions.
This and other related issues are presently under consideration.

\acknowledgments
The authors would like to thank C. Furtado for discussions,
and PROCAD/CAPES for financial support. DB thanks CNPq for partial support
and ARG thanks FAPEMA for a fellowship.

\end{document}